\documentclass[12pt]{article}

    \usepackage[utf8]{inputenc}
    \usepackage[
        twoside,
        top=1in,
        bottom=0.75in,
        inner=0.75in,
        outer=0.75in
    ]{geometry}
\usepackage[square, numbers]{natbib}
\usepackage{hyperref}
\usepackage{authblk}
\usepackage{graphicx}
\usepackage{moresize}
\usepackage{amsmath}
\usepackage{xcolor}
\usepackage{listings}
\usepackage{subcaption}
\newcommand{\match}[1]{\textbf{\textcolor{blue}{#1}}}

\definecolor{boxfillcolor}{HTML}{f1f1f1}
\definecolor{boxframecolor}{HTML}{ffffff}
\definecolor{promptdelim}{HTML}{4aa567}
\definecolor{background}{HTML}{f1f1f1}
\definecolor{delim}{HTML}{999999}
\definecolor{eclipseStrings}{HTML}{333333}
\definecolor{eclipseKeywords}{HTML}{444444}
\colorlet{numb}{gray!60!black}
\lstdefinelanguage{text}{
    basicstyle=\tiny  \ttfamily, %or \small or \footnotesize etc.
    stringstyle=\color{eclipseKeywords}, % style of strings
    breaklines=true,
    breakindent=0pt,
    framextopmargin=5pt,
    framexbottommargin=5pt,
    framexleftmargin=3pt,
    framexrightmargin=3pt,
    backgroundcolor=\color{background},
    literate=
        {\{}{{{\color{promptdelim}{\{}}}}{1}
        {\}}{{{\color{promptdelim}{\}}}}}{1},
}

    \title{ArchSeek: Retrieving Architectural Case Studies Using Vision-Language Models}
    \author[1]{Danrui Li}
    \author[2]{Yichao Shi}
    \author[3]{Yaluo Wang}
    \author[4]{Ziying Shi}
    \author[1,5]{Mubbasir Kapadia}
    \affil[1]{\small Rutgers University, NJ, USA}
    \affil[2]{\small Georgia Institute of Technology, GA, USA}
    \affil[3]{\small Harvard University, MA, USA}
    \affil[4]{\small Southeast University, Nanjing, China}
    \affil[5]{\small Roblox, CA, USA}
    \date{}

\begin{document}

%%%% Format Running Header %%%%%
\markboth{Danrui Li, Yichao Shi, Yaluo Wang et al.}{ArchSeek: Retrieving Architectural Case Studies Using Vision-Language Models}

%%%% Insert the Title Information %%%
\maketitle

%%%% General Description of the Document %%%%
\begin{abstract}
Efficiently searching for relevant case studies is critical in architectural design, as designers rely on precedent examples to guide or inspire their ongoing projects. However, traditional text-based search tools struggle to capture the inherently visual and complex nature of architectural knowledge, often leading to time-consuming and imprecise exploration. This paper introduces ArchSeek, an innovative case study search system with recommendation capability, tailored for architecture design professionals. Powered by the visual understanding capabilities from vision-language models and cross-modal embeddings, it enables text and image queries with fine-grained control, and interaction-based design case recommendations. It offers architects a more efficient, personalized way to discover design inspirations, with potential applications across other visually driven design fields. The source code is available at \url{https://github.com/danruili/ArchSeek}.

\end{abstract}

\section{Introduction}

The ability to efficiently locate and analyze relevant case studies is paramount in architectural design, where architects and designers often rely on online resources to find inspiration and precedent examples that align with their current projects~\cite{lee_how_2014,wu_reconfiguring_2025}. Online platforms like Google, Pinterest, or ArchDaily have become common tools for such exploration. However, these general-purpose search engines rely primarily on text-based search mechanisms~\cite{domeshek1992case} or noisy user interaction history, often requiring users to manually sift through vast collections of loosely related images or projects. This approach is time-consuming and frequently fails to capture the complex visual and contextual nature of architectural information.

This challenge stems from the fact that architectural design inherently involves spatial, stylistic, and material considerations that are difficult to fully describe or retrieve using textual queries or user interactions alone~\cite{GLASER2019141}. Recent advancements in Large Language Models (LLMs)~\cite{openai_gpt-4vision_2023} and joint learning across modalities~\cite{Cherti_2023} show promise in bridging this gap, as demonstrated in domains like digital storytelling~\cite{words2worldsLi2024} and game design~\cite{li2025cardiverse}. Nonetheless, their performance in architectural design remains unknown.

This paper presents ``ArchSeek", a novel system designed to augment the architectural case study process by enabling multimodal search and recommendation tailored to the domain’s unique needs. By augmenting architecture design case data with a vision-language model and indexing them with cross-modal embeddings, the system sets up a design case database with domain-specific knowledge, enabling the following capabilities:
\begin{enumerate}
    \item Natural language search with flexible granularity—ranging from simple keywords to detailed project descriptions.
    \item Image-based search image with fine-grained control over the attended perspectives.
    \item Implicit in-session recommendation based on user interactions, enabling discovery beyond explicit queries.
\end{enumerate}

While ArchSeek shares surface similarities with existing search engines, it differentiates itself by leveraging cross-modal embeddings and domain-specific data augmentation to offer a more personalized, context-aware experience. This allows architects to retrieve not just visually similar cases, but also those aligned with conceptual or spatial intentions—capabilities that general-purpose tools often lack. While the system performance has not reached an optimum, we believe it demonstrates its potential in architecture design practice and its generalizability to other design fields such as industrial design or graphic design.

\section{Related Works}

Traditional architectural design search systems primarily utilize text-based techniques, including keyword matching or predetermined tags, to collect case information. However, these techniques frequently struggle with efficiently handling the intricate image data associated with architectural design contexts~\cite{domeshek1992case}. Therefore, developments in architectural design search systems largely focus on image data. For example, Graph Neural Networks (GNNs) are employed in floor plans to identify architectural components based on design criteria. This can assist architects in the initial design phase~\cite{PARK2023106378}. Generative Adversarial Networks (GANs), on the other hand, enable dynamic adaptation of floor plans and real-time furniture arrangement~\cite{nauata2020houseganrelationalgenerativeadversarial}. Despite their primary focus on imagery, these methods struggle to effectively manage the complexity of architectural design projects that combine both image and text data.

In recent years, more personalized search experience, precisely described as recommendation systems, have evolved significantly. These systems can generate customized recommendations based on user preferences and historical data using techniques such as collaborative filtering and content analysis~\cite{palma2023}. The multilingual recipe recommendation platforms reflect these characteristics~\cite{Twomey_2020}. Advancements in large language models (LLMs) have allowed LLM-based recommender systems to produce more consistent and high-quality content~\cite{palma2023}. Users can dynamically improve search results through interaction with LLMs. Regarding image data, as illustrated in~\cite{wang2024techgpt20largelanguagemodel}, LLMs have the capability to process intricate visual data. Consequently, it has the capability to offer users tailored suggestions for graphic design projects.
  
When it comes to using different LLMs to make search or recommender systems better, Visual Language Models (VLMs) show a lot of promise in the field of architectural design because they combine visual and textual data in a way that makes it easier to understand complex architectural design cases. Vision Language Models (VLMs) such as GPT-4 Vision and ImageBind have shown strong multimodal performance in areas such as creative and general visual comprehension~\cite{Cherti_2023,openai_gpt-4vision_2023,zhang2023gpt4visiongeneralistevaluatorvisionlanguage}. Consequently, VLMs are competent in addressing complex visual and semantic relationships in architectural case studies, thus addressing the limitations in traditional architectural search systems.

\section{Methods}

\begin{figure}
\includegraphics[width=\textwidth]{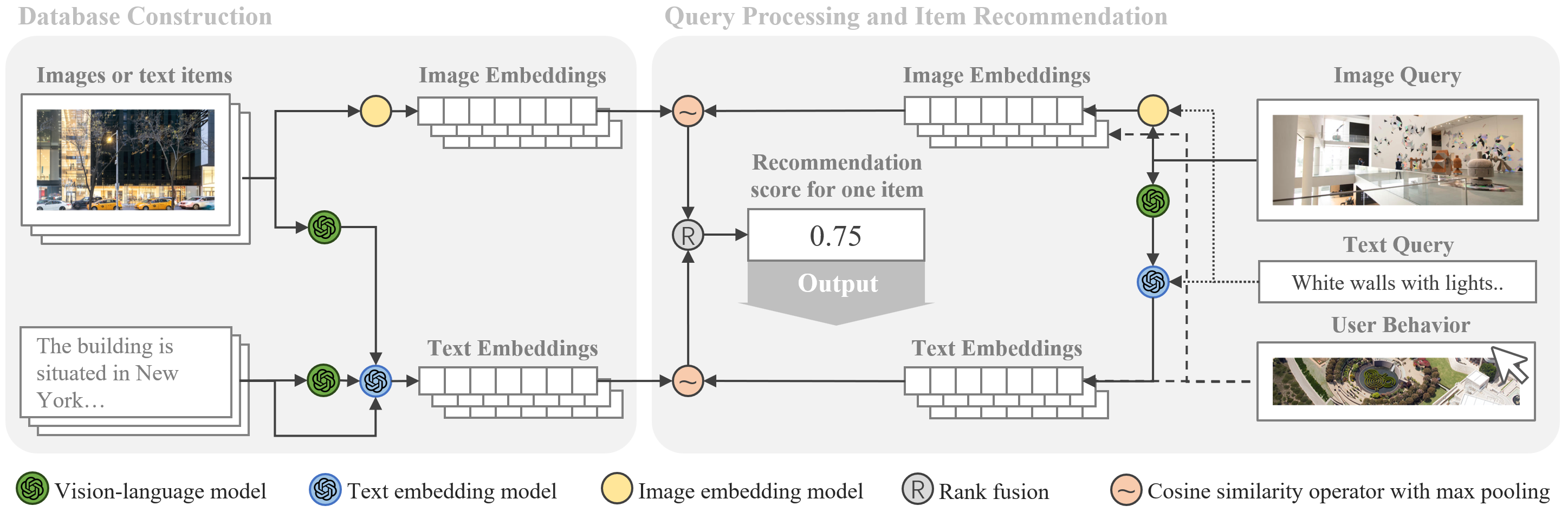}
\caption{The framework of ArchSeek shows the database construction stage and the query stage for single design case. In the database construction stage (left), all media files of the design case are augmented by a vision language model, generating architecture design reviews from various aspects. Then, their embeddings are generated for later use. In the query and recommendation stage (right), all three types of user interactions are converted into embeddings, compared to the embeddings of database items.} \label{framework}
\end{figure}

Our framework facilitates three user interaction modes (text query, image query, and automated recommendation), all of which operate through the following two components (Figure \ref{framework}):
\begin{itemize}
    \item Database Construction: We leverage state-of-the-art cross-modal models to generate a comprehensive database enriched with augmented textual analysis and cross-modal embeddings.
    \item Query Processing and Recommendation: The system employs a rank-based fusion methodology to integrate search results from both textual and visual comparisons. These comparisons are computed using cosine similarity metrics between the corresponding embedding vectors. In recommendation, the system augments queries with historical user interaction data, specifically focusing on 'Like' events, to facilitate autonomous design case recommendations.
\end{itemize}

\subsection{Survey-informed database construction}
\label{sec:database}

We constructed a dataset comprising 54 architectural design cases. For each case, we manually collected textual descriptions and images from online sources. The images encompass a balanced range of common categories, including bird's-eye views, ground-level photos, and architectural drawings.
Given limited resources, our study focuses on a specific subdomain within architectural design: newly constructed art galleries and museums worldwide.
However, we argue that our methodology can be generalized to a broader scope where multiple architectural categories exist.

\begin{figure}[ht]
\includegraphics[width=\textwidth]{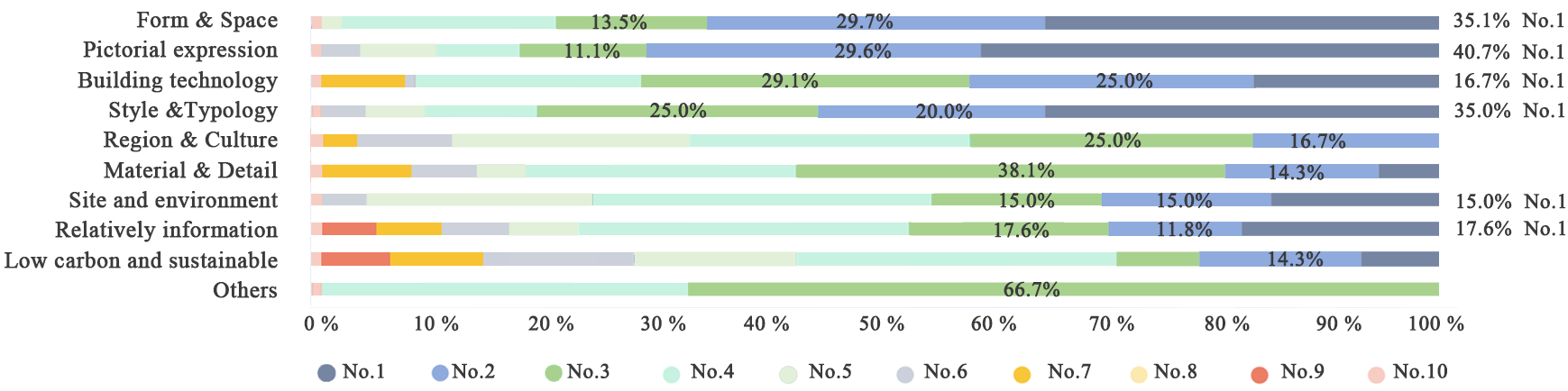}
\caption{User attention distribution on different topics of a design case when using architecture design case recommender systems.} \label{fig:survey}
\end{figure}

To align our database with the interests of our target users and their search behaviors in design practice, we conducted a survey involving architecture students (65\%) and professional architects (30\%), analyzing responses from 43 participants. The result (see Figure~\ref{fig:survey}) shows that users primarily focus on architectural forms, styles, expressions, and relationships with the environment.

\begin{figure}[ht]
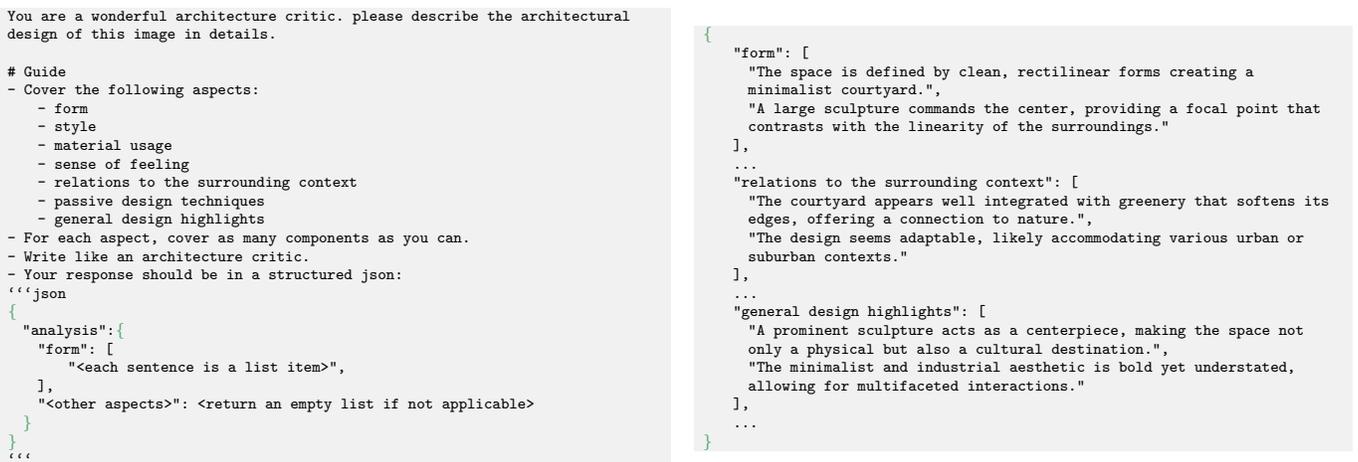

\begin{minipage}{.49\linewidth}
\begin{lstlisting}[language=text,numbers=none]
You are a wonderful architecture critic. please describe the architectural design of this image in details. 

# Guide
- Cover the following aspects: 
    - form
    - style
    - material usage
    - sense of feeling
    - relations to the surrounding context
    - passive design techniques
    - general design highlights
- For each aspect, cover as many components as you can.
- Write like an architecture critic. 
- Your response should be in a structured json:
```json
{
  "analysis":{
    "form": [
        "<each sentence is a list item>",
    ],
    "<other aspects>": <return an empty list if not applicable>
  }
}
```
\end{lstlisting}
\end{minipage}
\hfill
\begin{minipage}{.48\linewidth}
\begin{lstlisting}[language=text,numbers=none]
{
    "form": [
      "The space is defined by clean, rectilinear forms creating a minimalist courtyard.",
      "A large sculpture commands the center, providing a focal point that contrasts with the linearity of the surroundings."
    ],
    ...
    "relations to the surrounding context": [
      "The courtyard appears well integrated with greenery that softens its edges, offering a connection to nature.",
      "The design seems adaptable, likely accommodating various urban or suburban contexts."
    ],
    ...
    "general design highlights": [
      "A prominent sculpture acts as a centerpiece, making the space not only a physical but also a cultural destination.",
      "The minimalist and industrial aesthetic is bold yet understated, allowing for multifaceted interactions."
    ],
    ...
}
\end{lstlisting}
\end{minipage}
\caption{Using the vision-language model to extract analysis text from design case images. (left) The text prompt is used when calling the model. (right) A snippet of an output example.} \label{fig:prompt}
\end{figure}

In response to these topics, we employed a vision-language model (GPT-4-Vision~\cite{openai_gpt-4vision_2023}) to extract supplementary textual analyses from the images and text in our database. Each image or text file was inputted into the model with a prompt designed to elicit a critique akin to an architecture critic's. To ensure alignment with user interests, we crafted the prompts based on the main topics highlighted in the survey (see Figure~\ref{fig:prompt}). The results of the analysis were then stored for further use.

Each design case in the database was thus augmented by a collection of text entries derived from related images or text descriptions. These entries, along with segmented original text, were encoded into vector representations using OpenAI's “text-embedding-3-large” model [18] and Meta's ImageBind~\cite{girdhar2023imagebind}, which serve as compressed representations of the original text. Meanwhile, all images were embedded into vectors using ImageBind. The vectorized data was then stored for subsequent query processing, as elaborated in the following sections.

\subsection{User query and Item recommendation}

Our system provides two main query modes (Text Query and Image Query) as well as a recommendation mechanism based on user interactions. The following sections outline their usage through the graphical user interface (UI) and explain the underlying methodologies:

\subsubsection{Text Query mode} 
\begin{figure}
\centering
\includegraphics[width=\textwidth]{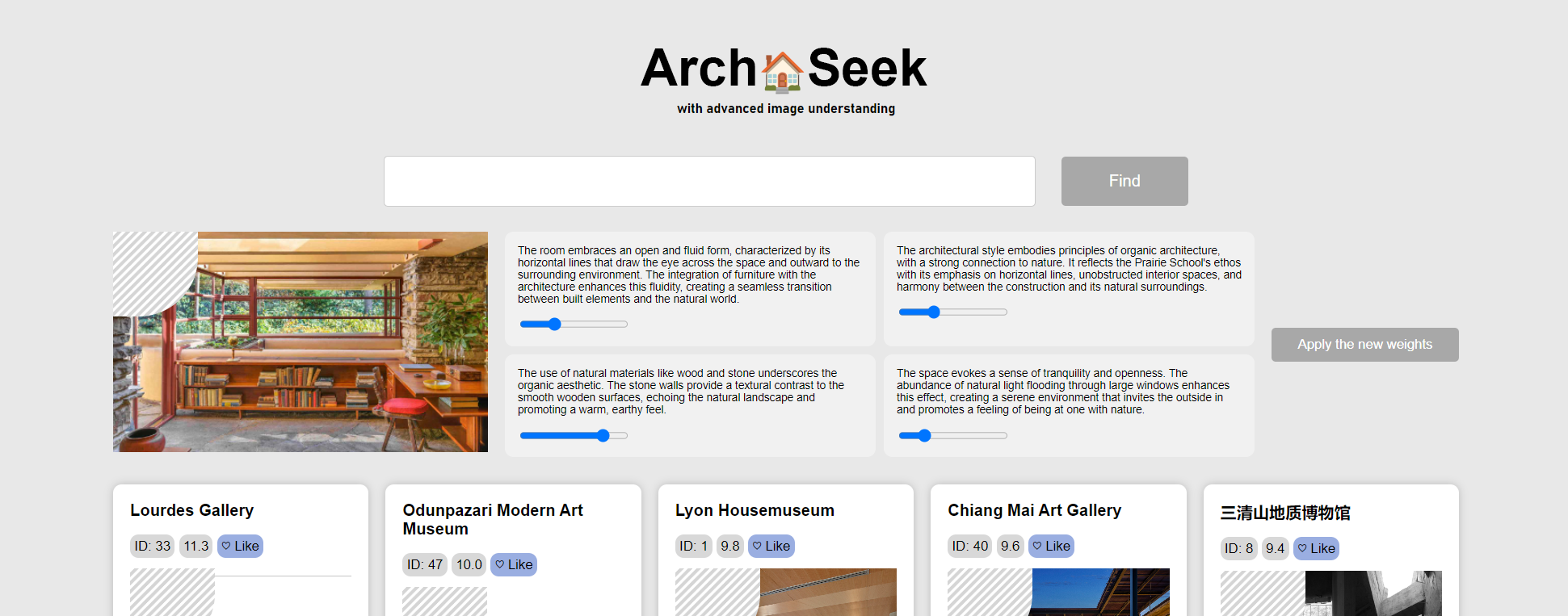}
\caption{The user interface of ArchSeek in Image Query mode. The interface displays image analysis and adjustable weight parameters via slider bars, followed by the retrieved design cases. The thumbnails of the design cases are partially masked for Fair Use Policy compliance.} \label{fig:ui}
\end{figure}

In this mode, users input design concepts to a text field in natural language, then click "Find" to search for relevant design cases from the database. The system ranks all design cases based on their relevance to the query, with each result linked to a detailed page for further exploration (see Figure \ref{fig:ui}).

The query integrates results from two search processes: text analysis and image understanding. In the text analysis search, the user's query $q$ is compared against all text entries $d$ for each design case $D$, including those extracted from images. The design cases are then ranked based on the relevance of their most closely matching text entry, determined by the cosine similarity of embedding vectors to measure the semantic distance between the query and database entries:

\begin{equation}
\text{Relevance}(q, D) = \max_{d \in D}(\frac{\text{Emb}(q) \cdot \text{Emb}(d) }{||\text{Emb}(q)||\cdot||\text{Emb}(d)||})
\end{equation}
where $\text{Emb}(\cdot)$ represents OpenAI’s text embedding model.

In the image understanding search, the comparison is conducted directly between the user query and the images (not their text analysis) in the database. This process follows the same method as above, except using ImageBind~\cite{girdhar2023imagebind} as the embedding model, which enables cross-modal comparison capabilities.

To combine the search results from text and image queries, the Reciprocal Rank Fusion (RRF) algorithm is used. It assigns a score to each rank in each search result, aggregating these scores across different lists. The results are then re-ranked based on the total scores. For a design case $D$ and query $q$, the RRF score calculation is expressed as:

\begin{equation}
\text{RRF}(q,D) = \frac{1}{\text{TextRank}(q,D) + c} + \frac{1}{\text{ImageRank}(q,D) + c}
\end{equation}
where $c$ is a parameter set as $10$ in this paper.

\subsubsection{Image Query Mode} This mode follows a similar process as text query but accounts for the complexity of image interpretation. In this mode, the system first analyzes the image input across various architectural topics before proceeding with retrieval. Users can adjust the relative importance of different topics—such as form, style, material, and emotional expression—via slider bars, allowing for customization of search results by emphasizing specific attributes.
 
When an image is used as the input, it undergoes analysis similar to that used for database images during pre-processing (see \ref{sec:database}). These analyses of generated text are considered as independent text queries, each of them producing a certain set of retrieval results. Results are ranked finally by linear combinations of scores across these sets, with the option of weight adjustment by a user in order to fine-tune the recommendations.

\subsubsection{In-session Recommendation} The recommendation system is activated when a user `likes' a design case by clicking the `like' button on its card. The system re-ranks the search results to prioritize the most relevant cases. Users have the option to `like' multiple cases, with the results updating dynamically over time. The recommendation can enhance results derived from either of the two query modes or initiate from a completely random set of results.

The recommendation mechanism is achieved by augmenting existing queries, using the text description from the user-liked design case. After users like some cases, the query process will involve multiple text queries which consist of the text descriptions from the liked cases and users' original input (if provided).

\subsection{Evaluation}
\label{sec:eval}

The evaluation process is conducted in three parts as follows:
\begin{itemize}
    \item Text Query: We begin by qualitatively analyzing how the system adapts to different user inputs; examples will be given to illustrate this. Further, we carry out a quantitative evaluation by using human-labeled test data, since the text query mode is the cornerstone of our system.
    \item Image Query and Recommendation: The image query mode of the system reveals its flexibility in how users can change the weights of different topics to personalize their search results. The recommendation mechanism gives examples that illustrate the process of item recommendation, highlighting how the system responds to user interactions and preferences.
    \item Comprehensive User Study: Finally, we conduct a user study to evaluate the system as a whole. This stage involves testing the system with participants to gather feedback and insights on overall usability, effectiveness, and user satisfaction.
\end{itemize}
We elaborate the protocol of quantitative evaluation in text query mode and user study as below.

\subsubsection{Quantitative evaluation}
We manually constructed an evaluation dataset comprising 77 query-item pairs (see first two columns in Tab. \ref{tab:gt}). Each pair consists of a query text expressed in various writing styles and perspectives, along with a corresponding list of relevant design cases (ground truth labels) within the database. The lists contain no ordering information.

\begin{table}
\caption{Query-item pair examples (first two columns) from the evaluation dataset and its corresponding top-5 retrieved results (last column) generated by our system. Successful matching IDs are displayed in bold text and highlighted with color. }\label{tab:gt}
\centering
\begin{tabular}{lll}
\hline
Query & Human-labeled IDs & Top-5 Retrieved \\
\hline
Glass facade with panoramic views &  9,\match{16},\match{36},\match{37}&  \match{16},\match{36},15,\match{37},4  \\
Architecture that evokes a sense of mystery &  15,\match{14},42,\match{39}&  8,11,10,\match{39},\match{14} \\
Symbiosis with water bodies & 3,5,7,\match{9},11,\match{14},15,26,34&  53,40,12,\match{14},\match{9} \\
maximize natural ventilation & \match{6},7,14,25&  47,21,\match{6},18,33 \\
Futuristic style & 2,\match{14},32&  \match{14},12,34,11,5 \\
\hline
\end{tabular}
\end{table}

The queries are input into our system during the evaluation, and the retrieved case labels (last column in Tab.~\ref{tab:gt}) are compared against the human-annotated ground truth labels (second column). 
The comparison quality is measured by top-k precision and recall:

\begin{equation}
\text{Precision@k} = \frac{|\{ \text{relevant items in top } k \}|}{k}
\end{equation}
\begin{equation}
\text{Recall@k} = \frac{|\{ \text{relevant items in top } k \}|}{|\{ \text{total relevant items} \}|}
\end{equation}
where top-k precision reflects the proportion of relevant items among the top-k retrievals, while recall measures the proportion of relevant items captured within the top-k retrievals.

We compare our proposed method against its ablated variants and baseline approaches to emphasize the advantages of incorporating visual information. Specifically, we evaluate two ablated versions: \texttt{No text augmentation} (red in Fig. \ref{fig:metric}), which omits the text augmentation process, and \texttt{No image embedding} (blue in Fig. \ref{fig:metric}), which excludes direct image embeddings generated using the ImageBind model.
For baseline comparisons, we include a text-only approach and a random baseline. In the text-only approach (green in Fig. \ref{fig:metric}), only the raw textual descriptions from the database are used as input, without utilizing GPT-Vision to extract supplementary information from images. Like our method, this approach generates text embeddings by segmenting descriptions into chunks and employs cosine similarity to identify the most relevant matches to a query. In the random baseline (gray in Fig. \ref{fig:metric}), the system disregards user input and returns a randomly sorted list of cases.

\subsubsection{User Study}

To evaluate the ArchSeek system's performance and user experience, we designed four tasks focusing on its core features: tag and category-based search, context-based relevance testing, personalized search with recommendation optimization, and image query. 

\begin{enumerate}
    \item Tag and category-based search: Participants will use short phrases, such as “modern architecture” or “eco-design”, to search related design cases.
    \item Context-based relevance testing: Participants will input a sentence or several sentences that contain multiple architectural design concepts to search cases.
    \item Personalized search and recommendation: Participants will interact with the system to optimize personalized recommendations by iteratively liking, disliking, and adjusting the weight of text prompts.
    \item Image query: Participants will upload a local image to search design cases based on the image content. 

\end{enumerate}

These tasks aimed to assess retrieval speed, result relevance, and user satisfaction, with users rating these aspects on a scale of 1 to 5.
In addition to task-based evaluations, we conducted interviews to gain deeper insights into user experiences and identify areas for improvement. The interview questions were designed to cover multiple dimensions ranging from overall operational efficiency, retrieval quality, limitations, and further suggestions.

\section{Results}

\subsection{Text Query} 
\begin{figure}
  \centering
\begin{subfigure}[t]{.85\linewidth}
\includegraphics[width=\linewidth]{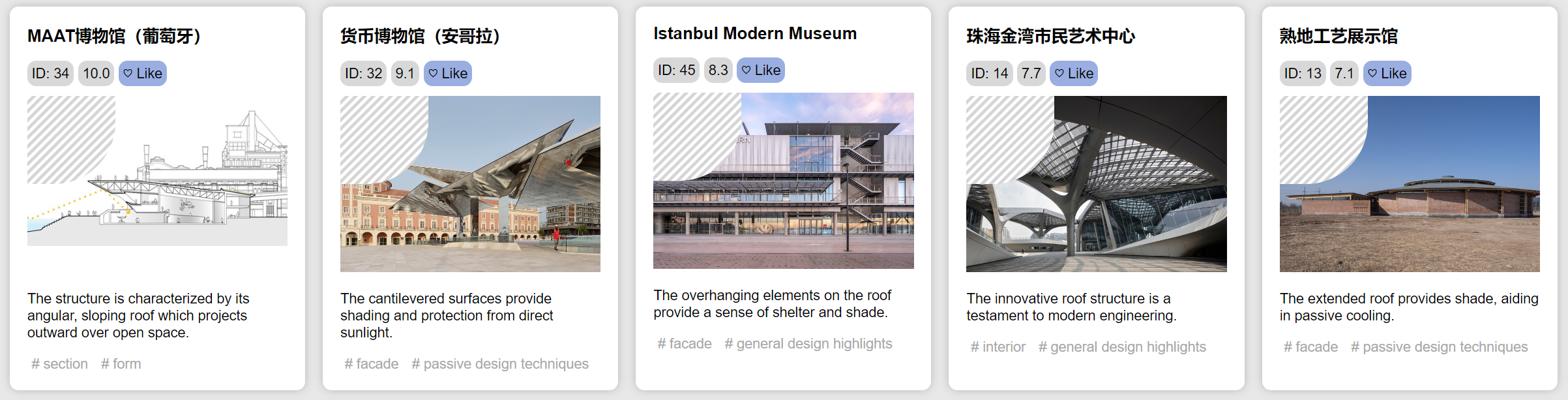}
\caption{Component query: cantilever roof}\label{fig:comp_query}
\end{subfigure}

\begin{subfigure}[t]{.85\linewidth}
\includegraphics[width=\linewidth]{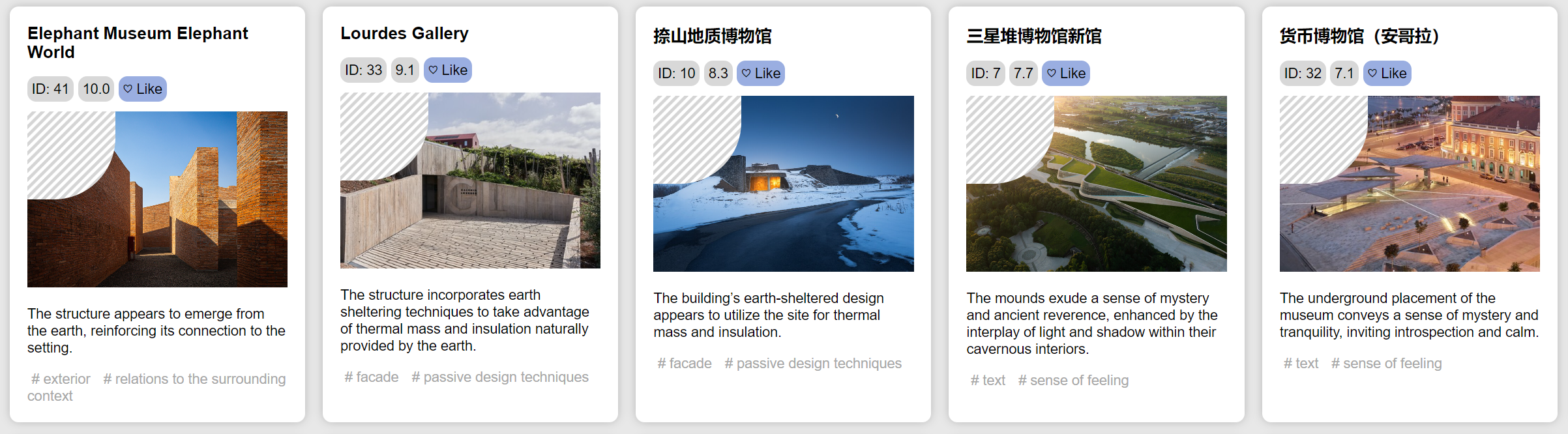}
\caption{Context query: buried in earth}\label{fig:context_query}
\end{subfigure}

\begin{subfigure}[t]{.85\linewidth}
\includegraphics[width=\linewidth]{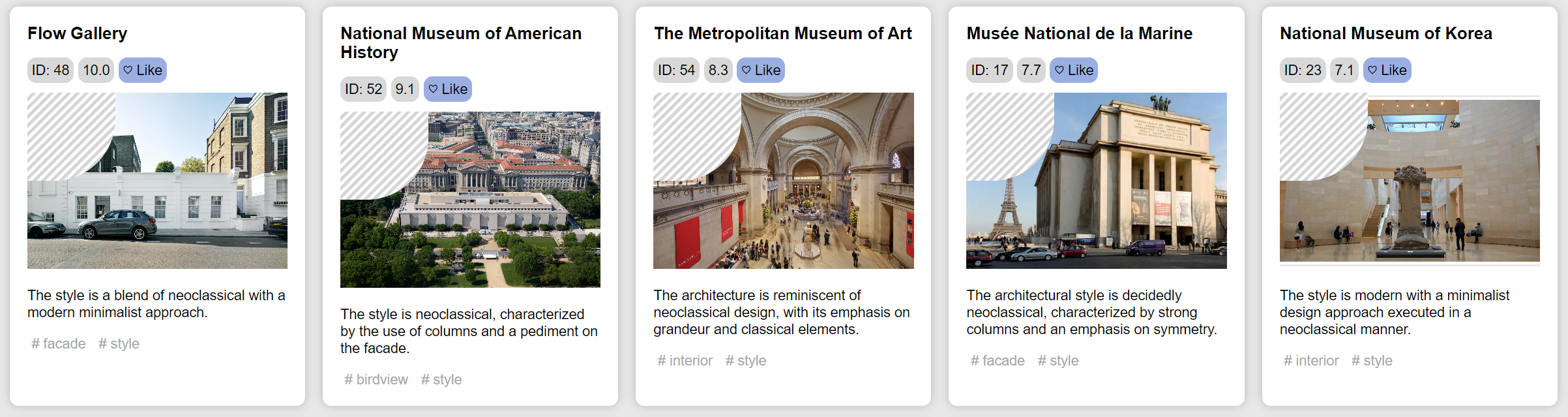}
\caption{Style query: Neoclassical style}\label{fig:style_query}
\end{subfigure}

\caption{Top five retrieved design cases using queries in various perspectives. Each retrieved design case comes with a similarity score (shown in light gray rounded rectangle below the title) and the most related description and image in the database. The thumbnails of the design cases are partially masked for Fair Use Policy compliance.} \label{fig:query}
\end{figure}

The qualitative results of text queries are illustrated in Figure \ref{fig:query}, where the top five retrieved design cases for different queries are presented. The queries vary in specificity, showcasing the system’s ability to handle diverse user inputs. These include essential architectural components (Figure \ref{fig:comp_query}), relationships with the surrounding environment (\ref{fig:context_query}), and stylistic elements (\ref{fig:style_query}). This capability addresses user needs identified in our survey (see Section 3.1), enabling more flexible and nuanced interactions with the search system.

\begin{figure}
\includegraphics[width=\textwidth]{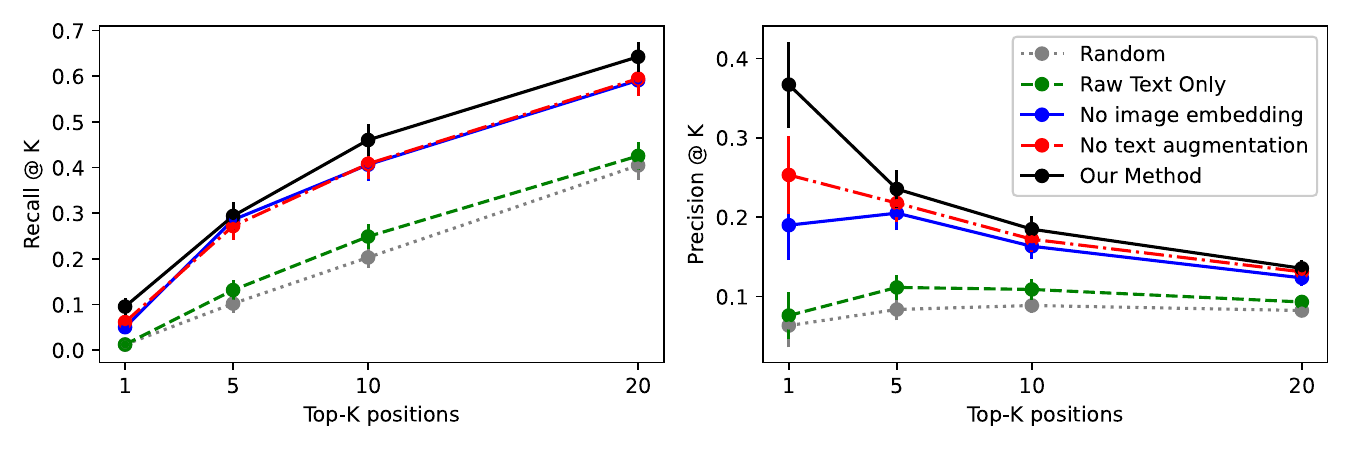}
\centering
\caption{The retrieval performance of ArchSeek on the evaluation dataset. The left plot shows the recall rates varying along the top-k position, while the right one shows the precision. The error bars represent the standard error of the mean values.} \label{fig:metric}
\end{figure}

The quantitative evaluation results, presented in Figure \ref{fig:metric}, demonstrate that our method achieves substantially higher performance metrics than both naive approaches and ablated variants in Section \ref{sec:eval}, although opportunities for further optimization in recall and precision remain. Notably, text-only methods, proven effective in other domains, show minimal advantages over random methods in our context. This under-performance can be attributed to the inherent characteristics of architectural design data, where visual content serves as the primary information carrier.

Furthermore, our approach outperforms methods relying solely on image embeddings without textual augmentation. We hypothesize that this superior performance stems from the limitations of general embedding models in capturing domain-specific architectural features.The integration of supplementary textual information guides the model toward attending to relevant architectural visual elements that might be overlooked. These findings emphasize the critical role of image comprehension capabilities in architectural design search systems, particularly the combination of image embeddings and textual augmentation.

% \vspace{-5mm}

\subsection{Image Query and Recommendation} 
\begin{figure}[ht]
\centering
\includegraphics[width=\textwidth]{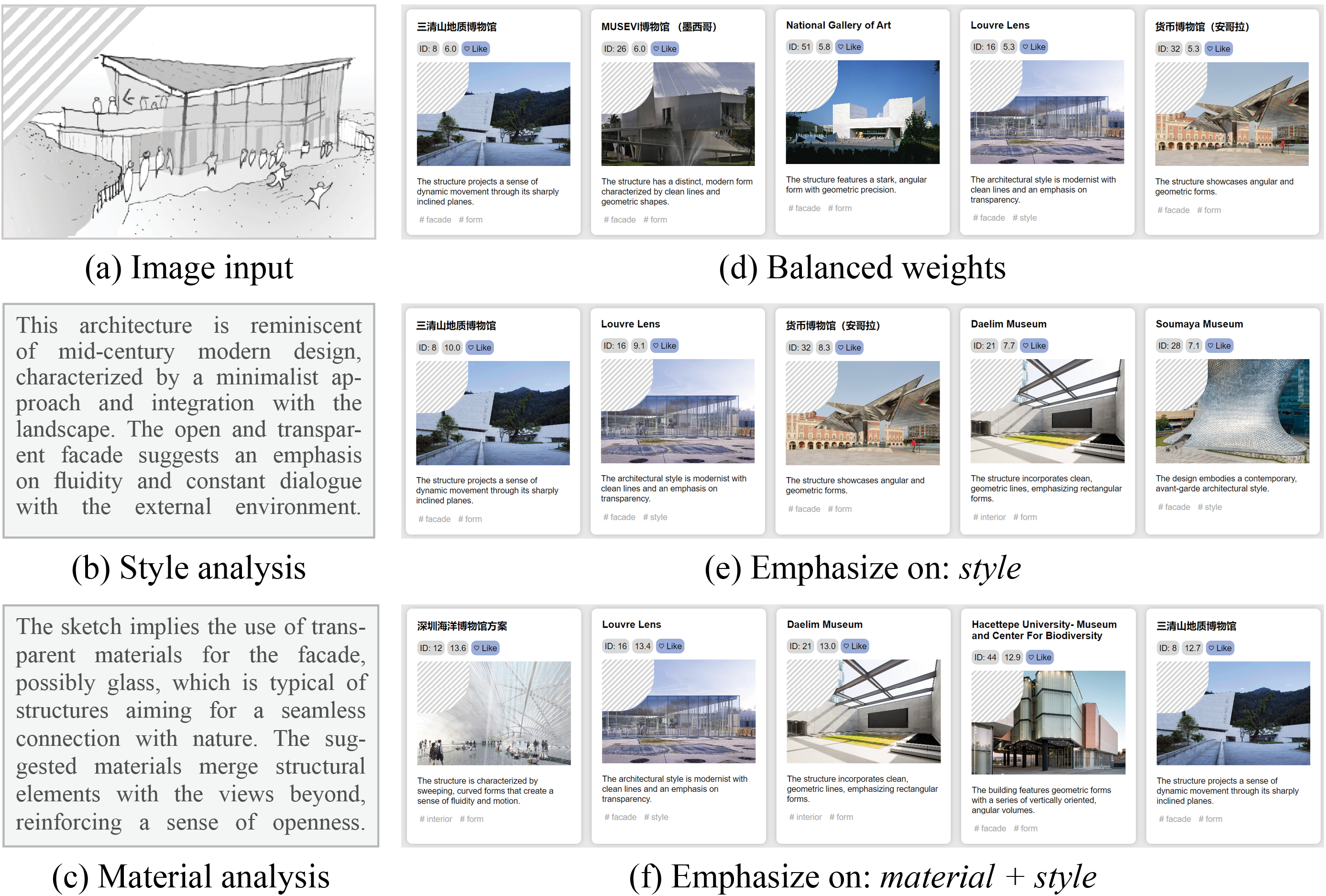}
\caption{Flexible design case search by adjusting emphasis during image input. The left column displays: (a) the example input image, (b) its corresponding style analysis, and (c) the material analysis. The right column (d-f) presents the top three retrieved design cases when different emphases are applied. The thumbnails of the design cases are partially masked for Fair Use Policy compliance.} \label{fig:adjust}
% \vspace{-4mm}
\end{figure}

Figure \ref{fig:adjust} showcases the flexibility of our system when handling image queries. First, it highlights the system's advanced image understanding capabilities, which accommodate a wide range of image styles. For example, an architectural drawing is used as input (Figure \ref{fig:adjust}a), and the system accurately interprets its content with sufficient inference regarding style and material usage (Figures \ref{fig:adjust}b and \ref{fig:adjust}c). Additionally, through the slider bars in the web app (see Figure \ref{fig:ui}), users can adjust the emphasis on different aspects, customizing the results according to their preferences. The adjusted results demonstrate the system’s responsiveness to user modifications (Figures \ref{fig:adjust}d and \ref{fig:adjust}e) and its capability to synthesize weights across various aspects (Figure \ref{fig:adjust}f).

\begin{figure}[h]

\begin{subfigure}[t]{\linewidth}
  \centering
\includegraphics[width=\linewidth]{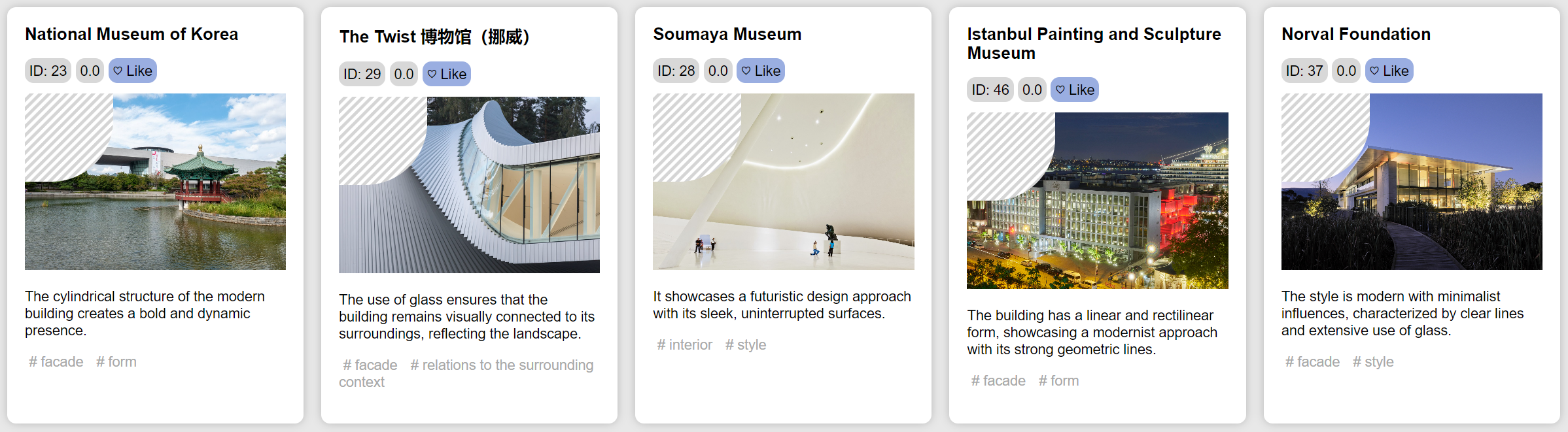}
\caption{Initiated Results. Only top five items are shown.} 
\label{fig:before-rec}
\end{subfigure}

\begin{subfigure}[t]{\linewidth}
  \centering
\includegraphics[width=\linewidth]{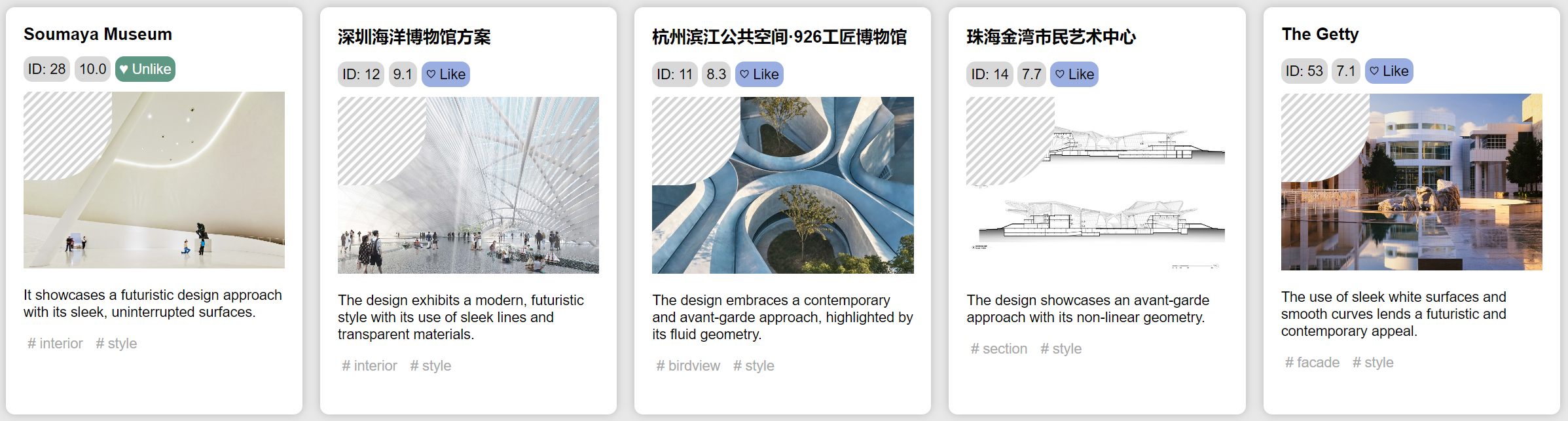}
\caption{Refreshed results after user picks one item.}
\label{fig:after-rec}
\end{subfigure}

\caption{\label{fig:rec} (top) The system is initialized with randomly recommended design cases. (bottom) After the user 'like' one item (the second one in the original order), the system refreshes the result with related design cases. The thumbnails of the design cases are partially masked for Fair Use Policy compliance.}\vspace{-2mm}
\end{figure}

Figure \ref{fig:rec} illustrates how the recommendations evolve as a user interacts with the system.
The system initially presents randomly selected recommendations to the user (Fig. \ref{fig:before-rec}). These are shown as a diverse set of design cases with varying attributes such as architectural styles, material use, and forms. Once the user interacts with the system by expressing a preference—specifically by “liking” the second item (ID:28, Soumaya Museum)—the recommendations are recalibrate.
The refreshed results (Fig. \ref{fig:after-rec}) show a significant shift, with the following notable patterns: 
(1) All refreshed designs focus on futuristic design that provides fluid forms and light color. (2) The interior and the style tags are dominated, just like the attributes emphasized in the selected user preference. Such a shift showcases the capability of our system in personalization, where results can converge on user preferences.

\subsection{User Study}

The user testing and interview process for the ArchSeek system involved assessing its core functionalities through four specific tasks: tag and category-based search, context-based retrieval, personalized search optimization, and retrieval via uploaded images.
Feedback (in Table \ref{tab:user}) revealed that simple retrieval tasks (Tasks 1 and 2) performed well in speed, though relevance and satisfaction were moderate, with some concerns about database diversity and result repetition. In contrast, personalized recommendations (Task 3) were praised for relevance and interactive features but faced occasional responsiveness issues. Image query (Task 4) exhibited quick speeds but struggled with precise matching and recognition of abstract content.

Interview feedback (Table \ref{tab:survey_summary}) further elaborated on system strengths and limitations. Users appreciated the system’s fast retrieval and interactive features, particularly in personalized recommendations, but emphasized the need for improved database diversity to reduce repetitive results and increase novelty. Tasks 2 and 4 highlighted challenges in matching results to complex queries or image content, pointing to the need for enhanced algorithmic precision. Suggestions included optimizing the interface, providing real-time feedback, and refining weight adjustment and navigation functionalities.

\begin{table}[ht]
\caption{User ratings on three metrics over four tasks. We report the average scores (scaled from 1 to 5) across 12 participants. }\label{tab:user}
\centering
\begin{tabular}{p{6cm}p{2cm}p{2cm}p{2cm}}
\hline
Task & Speed & Relevance & Satisfaction \\
\hline
Tag and category-based search& 3.75 & 3.42 &3.83\\
Context relevance testing & 3.75 & 3.58 &3.50\\
Personalized search and recommendation& 3.75 & 3.83 &4.00\\
Image query& 4.17 & 3.33 &3.33\\
\hline
\end{tabular}
\end{table}

% \vspace{-9mm}

\begin{table}[ht]

\caption{Summary of user interview}
\label{tab:survey_summary}
\centering
% \begin{tabular}{p{6cm}p{6cm}}
\begin{tabular}{p{.45\linewidth}p{.45\linewidth}}
\hline
 \textbf{Strengths} & \textbf{Issues / Suggestions} \\ 
\hline
\textit{Search and Database} & \\
\vspace{-15pt}
\begin{itemize}
    \item Fast search speed.
    \item Useful visual results.
    \item Keywords and labels provide some guidance.
\end{itemize}
\vspace{-15pt}& 
\vspace{-15pt}
\begin{itemize}
    % \item Results lack diversity, often repeat, and are not always relevant.
    \item Increase database size and diversity.
    \item Add filters and groups (e.g., by style, type, material).
\end{itemize} \vspace{-15pt}\\ \hline

\textit{Tags, Personalization, and Interaction} & \\
\vspace{-15pt}\begin{itemize}
    \item Versatile tags for filtering.
    \item Weight adjustments appreciated.
    \item Clear interface design.
\end{itemize}\vspace{-15pt} & 
\vspace{-15pt}\begin{itemize}
    % \item Provided labels are often too generic or irrelevant.
    \item ``Like" lowers relevance sometimes.
    \item Weight sliders are hard to use.
\end{itemize}\vspace{-15pt} \\ \hline

\textit{Special Features and General Feedback} & \\
\vspace{-15pt}\begin{itemize}
    \item Promising concept and features.
    \item Potential for creative inspiration.
\end{itemize}\vspace{-15pt} & 
\vspace{-15pt}\begin{itemize}
    \item Add preview windows when hovering before clicks.
    \item Multi-language support.
    % \item Expand and diversify database.
\end{itemize}\vspace{-15pt} \\ \hline
\end{tabular}

\end{table}

\section{Conclusion}
Traditional text-based search methods and recommendation systems fail to adequately represent the complicated interaction of visual and textual elements inherent in the construction of case data. This study has attempted to address these limitations. Based on large language models, ArchSeek, a case study search system with recommendation capability, is proposed in this paper. It comprises three modes of interaction: text query, image query, and dynamic recommendation. In each mode, the user can find applicable design examples through simple input or implicit "like" actions. This simplifies and personalizes the search process.

Initially, ArchSeek utilizes a search framework based on semantic similarity instead of keyword matching. Traditional methods depend on precise word matching, which does not adequately convey the specific significance of graphics utilized to illustrate architectural designs in specialized contexts. Utilizing LLM and visual language modeling, ArchSeek can extract content with related meanings. User evaluations indicate that this semantic search methodology greatly enhances search efficiency, eliminates redundant data exploration, and saves valuable time for users.

Secondly, ArchSeek offers a hybrid search framework that integrates both visual and textual architectural information. Architects typically use images to understand the aesthetics of a design, while they use text to delve into its technical and functional details. This search framework aligns better with the working practices of architects and the knowledge system attributes of the architectural design sector. Consequently, our system can proficiently satisfy architects' requirements for both visual and textual input.

Moreover, ArchSeek allows architects to effectively identify individual preferences, thereby personalizing the recommended results. This feature speeds up the workflow and decision-making processes of the architectural community. Moreover, clients utilizing architects' services may communicate their preferences more efficiently via the system, thereby enhancing collaboration between clients and architects.

Despite the beneficial aspects suggested by ArchSeek, the dataset's size is currently restricted by the focus primarily on art galleries and museums. We will expand the database in the future to encompass a wider range of architectural styles and building categories, thereby improving its adaptability. We will also develop stronger connections between this system and architectural design software. We will integrate ArchSeek with programs such as Rhino and Grasshopper. Furthermore, incorporating support for community collaboration and an AR/VR interface can improve the user experience by allowing users to get involved with the system in real time.

\bibliography{main}

\end{document}